# Faddeev-Senjanovic Quantization of Supersymmetrical Electrodynamical System

Yun-Guo JIANG[1]     Yong-Chang HUANG[1,2]

1. Institute of Theoretical Physics, Beijing University of Technology, Beijing 100022, China
2. CCAST ( World Lab. ), P. O. Box 8730, Beijing, 100080, China

## Abastract

According to the method of path integral quantization for the canonical constrained system in Faddeev-Senjanovic scheme, we quantize the supersymmetrical electrodynamic system in general situation, and obtain the generating functional of Green function. Another first class constraint is obtained by making the linear combination of several primary constraints, the generator of gauge transformation is constructed, gauge transformations of the all different fields are deduced. Utilizing the consistency equation of gauge fixing condition we deduce another gauge fixing condition, and we find that the secondary constraint of the system is an Euler-Lagrange equation that is just electro-charge conversation law. Thus, we do not need to calculate the other secondary constraints step by step, and get no new constraints naturally. So, the Faddeev-Senjanovic path integral quantization of the supersymmetrical electrodynamical system is simplified.

**Key Word**：supersymmetry, quantum electrodynamics, Faddeev-Senjanovic quantization, Dirac-Bergmman algorithm

## 1 Introduction

The minimal supersymmetrical standard model ( MSSM ) is currently the most favored candidate for extension of the standard model. Researches on supersymmetrical quantum field theories have great meaning to the possible discovery of supersymmetry.

Kushreshtha and Müller-Kirsten quantized 1+1 dimensional super fields in Faddevv-Jackiw scheme[1]; Batalin presented a superfield formulation of the quantization program for the theories with the first and second class constraints, and set up a phase-space path integral expression entirely in terms of superfields, further made BRST transformations and canonical transformations enter on equal footing[2]. Rupp et al obtained, in the level of supersymmetrical quantum field theory, Slazvnov-Taylor identity satisfying the invariance of the supersymmetrical transformations[3].

Supersymmetrical quantum electrodynamics (SQED) is a simple model of MSSM, and its action is the one of the singular Lagrange system. Using Faddeev-Senjanovic path integral quantization scheme we quantize the system of SQED, construct the generator of gauge transformation, and further give gauge transformation and the generating functional of Green functions.

## 2. Hamilton system

Under Wess-Zumino gauge, the SQED action is obtained in terms of its component fields after the integral of Grassmann coordinates of the action, i.e., the Lagrangian density is[4]

$$\mathcal{L}_{SQED} = -F_{\mu\nu}(x)F^{\mu\nu}(x) - 4i\lambda(x)\sigma^{\mu}\partial_{\mu}\overline{\lambda}(x) + i(D_{\mu}^{*}\overline{\psi_1}(x))\overline{\sigma}^{\mu}\psi_1(x)$$

$$+ i(D_{\mu}^{*}\overline{\psi_2})\overline{\sigma}^{\mu}\psi_2 + (D_{\mu}B_1)(D_{\mu}^{*}B_1^{*}) + (D_{\mu}B_2(x))(D_{\mu}^{*}B_2^{*}(x))$$

$$- \frac{e}{\sqrt{2}}\left[\overline{\lambda}(x)(\overline{\psi_1}(x)B_1(x) - \overline{\psi_2}(x)B_2(x)) + \lambda(x)(\psi_1(x)B_1^{*}(x) - \psi_2(x)B_2^{*}(x))\right]$$

$$+ m\left[\psi_1(x)\psi_2(x) + \overline{\psi_1}(x)\overline{\psi_2}(x)\right] - m^2\left[B_1(x)B_1^{*}(x) + B_2(x)B_2^{*}(x)\right]$$

$$- \frac{e^2}{32}\left[B_1(x)B_1^{*}(x) - B_2(x)B_2^{*}(x)\right], \tag{1}$$

where

$$D_{\mu} = \partial_{\mu} - \frac{1}{2}ieA_{\mu}, \quad D_{\mu}^{*} = \partial_{\mu} + \frac{1}{2}ieA_{\mu}, \quad F_{\mu\nu} = \partial_{\mu}A_{\nu} - \partial_{\nu}A_{\mu}, \tag{2}$$

$A_{\mu}$ is gauge boson field, $\lambda$ is the gaugino fermionic partner of $A_{\mu}$, $B_1$ and $B_2$ are supersymmetrical partners of $\psi_1$ and $\psi_2$ which are two-component fermionic matter fields, $F_{\mu\nu}$ $(\mu,\nu=0,1,2,3)$ are gauge field strength tensors, $e$ and $m$, respectively, represent electro-charge and mass parameters, All spinors are two-component Weyl spinors. We take metrics[5] to be $g_{\mu\upsilon} = (1,-1,-1,-1)$, and introduce the Pauli matrices $\sigma^0 = \begin{pmatrix} 1 & 0 \\ 0 & 1 \end{pmatrix}$,

$\sigma^1 = \begin{pmatrix} 0 & 1 \\ 1 & 0 \end{pmatrix}$, $\sigma^2 = \begin{pmatrix} 0 & -i \\ -i & 0 \end{pmatrix}$, $\sigma^3 = \begin{pmatrix} 1 & 0 \\ 0 & -1 \end{pmatrix}$, $\overline{\sigma}^0 = \sigma^0$, $\overline{\sigma}^i = -\sigma^i$.

## 3 Analysis for constraints of the Hamilton system

Since the Lagrangian of the SQED system is singular, we discuss its constraints in phase space, the canonical momenta conjugate to the component fields are

$$\pi_{\mu} = 4F_{\mu 0}, \qquad \pi_{\lambda} = 0, \qquad \pi_{\overline{\lambda}} = -4i\lambda\sigma^0,$$

$$\pi_{B_1} = D_0^* B_1^* = \left(\partial_0 + \frac{1}{2}ieA_0\right)B_1^*, \quad \pi^*_{B_1} = D_0 B_1 = \left(\partial_0 - \frac{1}{2}ieA_0\right)B_1,$$

$$\pi_{B_2} = D_0^* B_2^* = \left(\partial_0 + \frac{1}{2}ieA_0\right)B_2^*, \quad \pi^*_{B_2} = D_0 B_2 = \left(\partial_0 - \frac{1}{2}ieA_0\right)B_2,$$

$$\pi_{\psi_1} = 0, \quad \pi_{\overline{\psi}_1} = -i\overline{\sigma}^0 \psi_1, \quad \pi_{\psi_2} = 0, \quad \pi_{\overline{\psi}_2} = -i\overline{\sigma}^0 \psi_2. \tag{3}$$

The canonical Hamiltonian density is given by

$$\mathcal{H}_c = -A_0 \partial_i \pi^i - \frac{1}{8}\pi_i \pi^i + \pi_{B_1}\pi_{B_1^*} + \frac{1}{2}ieA_0\left(\pi_{B_1}B_1 - \pi_{B_1^*}B_1^*\right) + \pi_{B_2}\pi_{B_2^*} + \frac{1}{2}ieA_0\left(\pi_{B_2}B_2 - \pi_{B_2^*}B_2^*\right)$$

$$+\frac{1}{2}eA_0\left(\overline{\psi}_1\overline{\sigma}^0\psi_1 + \overline{\psi}_2\overline{\sigma}^0\psi_2\right) + F_{ij}F^{ij} - (D_i B_1)(D_i^* B_1^*) - (D_i B_2)(D_i^* B_2^*) - i(D_i^*\overline{\psi}_1)\overline{\sigma}^i\psi_1$$

$$-i(D_i^*\overline{\psi}_2)\overline{\sigma}^i\psi_2 + \frac{e}{\sqrt{2}}\left[\overline{\lambda}\left(\overline{\psi}_1 B_1 - \overline{\psi}_2 B_2\right) + \lambda\left(\psi_1 B_1^* - \psi_2 B_2^*\right)\right] + 4i\lambda\sigma^k\partial_k\overline{\lambda} - m\left(\psi_1\psi_2 + \overline{\psi}_1\overline{\psi}_2\right)$$

$$-m^2\left(B_1 B_1^* + B_2 B_2^*\right) + \frac{e^2}{32}\left(B_1 B_1^* - B_2 B_2^*\right). \tag{4}$$

According to Dirac constraint theory[6], it follows that there are 7 primary constraints

$$\phi_1^0 = \pi_0 \approx 0, \quad \phi_2^0 = \pi_\lambda \approx 0, \quad \phi_3^0 = \pi_{\overline{\lambda}} + 4i\lambda\sigma^0 \approx 0,$$

$$\phi_4^0 = \pi_{\psi_1} \approx 0, \quad \phi_5^0 = \pi_{\overline{\psi}_1} + i\overline{\sigma}^0\psi_1 \approx 0, \quad \phi_6^0 = \pi_{\psi_2} \approx 0,$$

$$\phi_7^0 = \pi_{\overline{\psi}_2} + i\overline{\sigma}^0\psi_2 \approx 0, \tag{5}$$

where the symbol " $\approx$ " means weak equality in Dirac sense [7].

The total Hamiltonian is given by

$$H_T = \int d^4 x \left(\mathcal{H}_c + u_1\phi_1^0 + u_2\phi_2^0 + u_3\phi_3^0 + u_4\phi_4^0 + u_5\phi_5^0 + u_6\phi_6^0 + u_7\phi_7^0\right). \tag{6}$$

The consistency equations of primary constraints are

$$\dot{\phi}_l^0 = \{\phi_l^0, H_T\}_P \approx 0, \quad (l = 1,2,3,4,5,6,7,). \tag{7}$$

Assume that F and G are functions of the Grassmann canonical variables ($\eta^\alpha, \pi_\alpha$), the Possion Bracket is given by[8]

$$\{F,G\} = \frac{\partial_r F}{\partial \eta^\alpha}\frac{\partial_l G}{\partial \pi_\alpha} - (-1)^{n_F n_G}\frac{\partial_r G}{\partial \eta^\alpha}\frac{\partial_l F}{\partial \pi_\alpha}, \tag{8}$$

where $n_F, n_G$ represent the Grassmann parities of functions F and G, respectively. The Lagrangian multipliers $u_2, u_3, u_4, u_5, u_6, u_7$ are solved out by consistency equations of primary constraints $\phi_l$ $(l = 2,3,4,5,6,7,)$. Consistency of $\phi_1$ leads to a secondary constraint

$$\phi_8^1 = \{\pi_0, H_T\}_P$$
$$= \partial_i \pi^i + \frac{1}{2} ie\left(-B_1 \pi_{B_1} - B_2 \pi_{B_2} + B_1^* \pi_{B_1^*} + B_2^* \pi_{B_2^*}\right) - \frac{1}{2} e\left(\overline{\psi_1} \overline{\sigma}^0 \psi_1 + \overline{\psi_2} \overline{\sigma}^0 \psi_2\right) \approx 0. \quad (9)$$

According to Dirac-Bergmman algorithm[8], three situations may occur from the consistency equations: (1) on the constraint's surface, we get an identity $0 = 0$; (2) the consistency equations are independent of Lagrangian multipliers, we get new constraints; (3) we get equations of Lagrangian multipliers. We get $\phi_8^1$ as a secondary constraint, it expresses the charge conversation law of the SQED system in phase space, and is just an Euler-Lagrange equation when converted to configuration space, which cannot give new constraint. When substituting the solved Lagrangian multipliers into the consistency equation, we also get an identity $0 = 0$, and cannot obtain new constraint.

We obtain another first-class constraint by making the linear combination of $\phi_4^0, \phi_5^0, \phi_6^0, \phi_7^0, \phi_8^1$, and further renew to mark the constraints as follows

$$\Lambda_1 = \phi_1^0,$$

$$\Lambda_2 = (\phi_8^1)' = \phi_8^1 - \frac{1}{2} ie\left(\overline{\psi_2}\theta_6 - \overline{\psi_1}\theta_4 - \psi_1\theta_3 + \psi_2\theta_5\right)$$
$$= \partial_i \pi^i + \frac{1}{2} ie\left(-B_1 \pi_{B_1} - B_2 \pi_{B_2} + B_1^* \pi_{B_1^*} + B_2^* \pi_{B_2^*}\right) + \frac{1}{2} ie\left(-\overline{\psi_2}\pi_{\overline{\psi_2}} - \overline{\psi_1}\pi_{\overline{\psi_1}} + \psi_1\pi_{\psi_1} + \psi_2\pi_{\psi_2}\right),$$

$$\theta_1 = \phi_2^0, \theta_2 = \phi_3^0, \theta_3 = \phi_4^0, \theta_4 = \phi_5^0, \theta_5 = \phi_6^0, \theta_6 = \phi_7^0, . \quad (10)$$

In terms of the definitions of Dirac's first and second-classe constraints[9], we obtain that $\Lambda_1, \Lambda_2$ are the first class of constraints, and $\theta_1, \theta_2, \theta_3, \theta_4, \theta_5, \theta_6$ are the second-class constraints. According to Castellani's method to construct generator of gauge transformation[10], we get the generator of the system as follows

$$G = \int d^3x[\dot{\varepsilon}(x)\Lambda_1 - \varepsilon(x)\Lambda_2]$$
$$= \int d^3x\{\dot{\varepsilon}(x)\pi_0 - \varepsilon(x)[\partial_i \pi^i + \frac{1}{2} ie\left(-B_1 \pi_{B_1} - B_2 \pi_{B_2} + B_1^* \pi^*_{B_1} + B_2^* \pi^*_{B_2}\right)$$
$$+ \frac{1}{2} ie\left(\overline{\psi_2}\pi_{\overline{\psi_2}} - \overline{\psi_1}\pi_{\overline{\psi_1}} + \psi_1\pi_{\psi_1} - \psi_2\pi_{\psi_2}\right)]\} . \quad (11)$$

Therefore, the transformations of the component fields are

$$\delta A_\mu = \{A_\mu(x), G\}_P = \partial_\mu \varepsilon(x), \qquad \delta\lambda = \{\lambda(x), G\} = 0,$$
$$\delta\overline{\lambda} = \{\overline{\lambda}(x), G\} = 0, \qquad \delta B_1 = \{B_1(x), G\} = \frac{1}{2} ie\varepsilon(x) B_1,$$
$$\delta B_1^* = \{B_1^*(x), G\} = -\frac{1}{2} ie\varepsilon(x) B_1^*, \quad \delta B_2 = \{B_2(x), G\} = \frac{1}{2} ie\varepsilon(x) B_2,$$
$$\delta B_2^* = \{B_2^*(x), G\} = -\frac{1}{2} ie\varepsilon(x) B_2^* \qquad \delta\overline{\psi_1} = \{\overline{\psi_1}(x), G\} = -\frac{1}{2} ie\varepsilon(x)\overline{\psi_1},$$

$$\delta\overline{\psi}_2 = \{\overline{\psi}_2(x), G\} = -\frac{1}{2}ie\varepsilon(x)\overline{\psi}_2, \quad \delta\psi_1 = \{\psi_1(x), G\} = \frac{1}{2}ie\varepsilon(x)\psi_1,$$

$$\delta\psi_2 = \{\psi_2(x), G\} = \frac{1}{2}ie\varepsilon(x)\psi_2, \quad \delta\pi_\mu = \{\pi_\mu(x), G\} = 0,$$

$$\delta\pi_\lambda = \{\pi_\lambda, G\} = 0, \quad \delta\pi_{\overline{\lambda}} = \{\pi_{\overline{\lambda}}(x), G\} = 0,$$

$$\delta\pi_{B_1} = \{\pi_{B_1}(x), G\} = -\frac{1}{2}ie\varepsilon(x)\pi_{B_1}, \quad \delta\pi_{B_2} = \{\pi_{B_2}(x), G\} = -\frac{1}{2}ie\varepsilon(x)\pi_{B_2},$$

$$\delta\pi^*_{B_1} = \{\pi^*_{B_1}(x), G\} = \frac{1}{2}ie\varepsilon(x)\pi^*_{B_1}, \quad \delta\pi^*_{B_2} = \{\pi^*_{B_2}(x), G\} = \frac{1}{2}ie\varepsilon(x)\pi^*_{B_2},$$

$$\delta\pi_{\overline{\psi}_1} = \{\pi_{\overline{\psi}_1}(x), G\} = \frac{1}{2}ie\varepsilon(x)\pi_{\overline{\psi}_1}, \quad \delta\pi_{\overline{\psi}_2} = \{\pi_{\overline{\psi}_2}(x), G\} = -\frac{1}{2}ie\varepsilon(x)\pi_{\overline{\psi}_2},$$

$$\delta\pi_{\psi_1} = \{\pi_{\psi_1}(x), G\} = -\frac{1}{2}ie\varepsilon(x)\pi_{\psi_1}, \quad \delta\pi_{\psi_2} = \{\pi_{\psi_2}(x), G\} = \frac{1}{2}ie\varepsilon(x)\pi_{\psi_2}. \quad (12)$$

These transformations are gauge transformations in phase space for the system.

## 4 Generating functional of Green function

The Lagrangian density is unchanged under the gauge transformations (12). According to path integral quantization in Faddeev-Senjanovic scheme[11], for each first-class constraint, we need to choose a gauge fixing condition. Consider the Coulomb gauge

$$\Omega_2 = \partial_i A^i \approx 0, \quad (i = 1, 2, 3). \tag{13}$$

Using the consistency of $\Omega_2$, we obtain another gauge fixing condition as follows

$$\Omega_1 = \nabla^2 A_0 - \frac{1}{4}\partial_i \pi^i \approx 0. \tag{14}$$

We introduce exterior sources of fields and their conjugate momenta $(\varphi_\alpha, \pi^\alpha)$, the generating functional of Green function for this system is given by

$$Z[J, K] = \int \mathcal{D}\varphi^\alpha \mathcal{D}\pi_\alpha \prod_{i,k,l} \delta(\Lambda_i)\delta(\Omega_k)\delta(\theta_l) \det|\{\Lambda_i, \Omega_k\}| \cdot (\det|\{\theta_{l_1}, \theta_{l_2}\}|)^{1/2}$$

$$\cdot \exp\{i \int d^4x (\pi_\alpha \dot{\varphi}^\alpha - \mathcal{H}_e + J_\alpha \varphi^\alpha + K^\alpha \pi_\alpha)\}. \tag{15}$$

It is easy to check that $\det|\{\Lambda_i, \Omega_k\}|$、$\det|\{\theta_{l_1}, \theta_{l_2}\}|$ are independent of the fields, and thus we can omit them from the generating functional of Green function, then we have

$$Z[J, K] = \int \mathcal{D}\varphi^\alpha \mathcal{D}\pi_\alpha \prod_{i,k,l} \delta(\Lambda_i)\delta(\Omega_k)\delta(\theta_l) \exp\{i \int d^4x (\mathcal{L}_{eff} + J_\alpha \varphi^\alpha + K^\alpha \pi_\alpha)\}. \tag{16}$$

Using the property of $\delta$ function

$$\delta(\Lambda_\alpha) = \int \frac{du^\alpha}{2\pi} \exp(iu^\alpha \Lambda_\alpha). \tag{17}$$

The generating functional of Green function for this system is now deduced as follows

$$Z[J,K] = \int \mathcal{D}\varphi^\alpha \mathcal{D}\pi_\alpha \mathcal{D}u_i \mathcal{D}v_j \mathcal{D}w_k \exp\{i\int d^4x(\mathcal{L}_{eff} + J_\alpha \varphi^\alpha + K^\alpha \pi_\alpha)\}, \qquad (18)$$

where

$$\mathcal{L}_{eff} = \mathcal{L}^p + \mathcal{L}_m, \qquad (19)$$

$$\mathcal{L}^P = \pi_a \varphi^a - \mathcal{H}_c, \qquad (20)$$

$$\mathcal{L}_m = u_i \Lambda_i + v_j \Omega_j + \omega_k \theta_k, \qquad (21)$$

$$\varphi^\alpha = \left(A_\mu, \lambda, \bar{\lambda}, B_1, B_2, B_1^*, B_2^*, \psi_1, \psi_2, \overline{\psi_1}, \overline{\psi_2}, u_i, v_j, w_k\right), \qquad (22)$$

$$\pi_\alpha = \left(\pi_\mu, \pi_\lambda, \pi_{\bar{\lambda}}, \pi_{B_1}, \pi_{B_1^*}, \pi_{B_2}, \pi_{B_2^*}, \pi_{\psi_1}, \pi_{\overline{\psi_1}}, \pi_{\psi_2}, \pi_{\overline{\psi_2}}\right), \qquad (23)$$

$$J_\alpha = \left(J_\mu, J_\lambda, J_{\bar{\lambda}}, J_{B_1}, J_{B_2}, J_{B_1^*}, J_{B_2^*}, J_{\psi_1}, J_{\psi_2}, J_{\overline{\psi_1}}, J_{\overline{\psi_2}}, J_{u_i}, J_{v_i}, J_{w_i}\right), \qquad (24)$$

$$K^\beta = \left(K_\mu, K_\lambda, K_{\bar{\lambda}}, K_{B_1}, K_{B_2}, K_{B_1^*}, K_{B_2^*}, K_{\psi_1}, K_{\psi_2}, K_{\overline{\psi_1}}, K_{\overline{\psi_2}}\right), \qquad (25)$$

where $u_i, v_j, \omega_k$ are the multiplier fields, and exterior sources $J_{u_i}, J_{v_j}, J_{w_k}$ corresponding to the multiplier fields are induced.

## 5 Summary and conclusion

Based on the constrained Hamilton theory, we obtain the constraints in the singular SQED system in phase space, two first-class and six secondary constraints are obtained through combining the primary and secondary constraints. Using spinor electrodynamics, Ref.[12] rigorously proved that the secondary constraints act as independent generators of gauge transformations (Dirac conjecture) for the system possessing only the first-class constraints, spinor field $\psi$ has the conjugate momentum $\pi_\psi = i\bar{\psi}\gamma^0$, but it is not considered as a constraint, and $\pi_{\bar{\psi}}$ is not introduced as the conjugate momentum of $\bar{\psi}$ in Ref.[12]. We find that the secondary constraint (9) is the electric charge conversation law of supersymmetry spinor electrodynamics. On the other hand, we may also use Faddeev-Jackiw quantization method[13] to quantize the supersymmetrical electrodynamic system.

According to Castellani's method to construct generator of gauge transformations[10], we get the generator of the gauge transformtions, and the gauge transformations of the component fields.

Using path integral quantization for canonical constrained system in Faddeev-Senjanovic scheme, and considering Coulomb gauge and its consistent equation to fix gauge, we quantize the supersymmetrical electrodynamic system, and get the generating functional of Green function for this system. Furthermore, we can obtain the canonical Ward identities for the system with the generating functional of Green function

**References**：